\documentclass[a4paper,11pt]{article}
\pdfoutput=1
\usepackage{jheppub}
\usepackage{textcase}
\usepackage{graphicx,epsfig,psfrag,amssymb,hyperref}
\usepackage{multirow}
\usepackage{color,graphicx,epsfig,psfrag,amsmath,empheq}
\usepackage[dvipsnames]{xcolor}
\usepackage{bm}
\usepackage{mathrsfs,amsfonts,soul,color}
\usepackage[caption=false]{subfig}
\usepackage{hepunits}
\usepackage{enumitem}
\usepackage{placeins}
\usepackage{textcomp}

\usepackage{float}
\usepackage[toc,page]{appendix}
\usepackage[ruled,vlined]{algorithm2e}
\usepackage{wrapfig}
\SetKwInOut{Input}{Input}
\SetKwInOut{Output}{Output}
\SetKwFunction{Fforward}{Forward}
\SetKwFunction{Fbackward}{Backward}
\SetKwProg{Fn}{function}{:}{}
\DontPrintSemicolon

\begin{document}

\title{Anomaly Detection for Physics Analysis \\ and Less than Supervised Learning} 

\author{Benjamin Nachman}

\affiliation{
\begin{scriptsize}
\phantom{ }\hspace{-0.12in}Physics Division, Lawrence Berkeley National Laboratory, Berkeley, CA 94720, USA \\
\phantom{ }\hspace{-0.12in}Berkeley Institute for Data Science, University of California, Berkeley, CA 94720, USA \\
\end{scriptsize}
}

\emailAdd{bpnachman@lbl.gov}

\abstract{
Modern machine learning tools offer exciting possibilities to qualitatively change the paradigm for new particle searches.   In particular, new methods can broaden the search program by gaining sensitivity to unforeseen scenarios by learning directly from data.  There has been a significant growth in new ideas and they are just starting to be applied to experimental data.  This chapter introduces these new anomaly detection methods, which range from fully supervised algorithms to unsupervised, and include weakly supervised methods.
}

\maketitle

\section{Introduction}
\label{sec:anomalies:intro}

Searching for new particles and forces of nature is one of the main goals of High Energy Physics (HEP).  Despite hints for new fundamental structure, there has been no convincing evidence since the discovery of the Higgs Boson in 2012~\cite{Chatrchyan:2012ufa,Aad:2012tfa}.   All of the major HEP experiments are engaged in an extensive search program and the goal of this chapter is to explore the dependence of these efforts on particular models and to examine how machine learning may be able to significantly broaden existing efforts.

In particular, the main topic of this chapter is machine learning-based \textit{anomaly detection}.  An anomaly is an unexpected feature of the data and as such all searches for new particles are anomaly detection analyses.  The main feature of the searches described in this chapter that separates them from others is the level of model dependence in various parts of the analysis.  In particular, this chapter will focus on searches that rely as little as possible on signal and background models for both achieving sensitivity to new physics as well as calibrating the background. This is in contrast to most dedicated searches that are optimized with a particular signal model.   All of these concepts will be made more precise below.

This chapter is organized as follows.  Section~\ref{sec:anomalies:model} introduces traditional searches for new particles in HEP and how model dependence plays a role in the analysis development and statistical procedure.  The remaining sections review various types of machine learning-based anomaly detection approaches.  These sections are organized by how labeled examples are (or are not) provided to train machine learning classifiers (\textit{supervision}).  In particular, the chapter covers supervised learning (Sec.~\ref{sec:fullysupervised} and Sec.~\ref{sec:supervised}), unsupervised learning (Sec.~\ref{sec:unsupervised}), weakly supervised learning (Sec.~\ref{sec:weaklysupervised}), and lastly hybrid approaches (Sec.~\ref{sec:hybrid}).  Recent results from ATLAS and CMS are highlighted in Sec.~\ref{sec:results} and the chapter ends with conclusions and outlook in Sec.~\ref{sec:outlook}.

\section{Model Dependence in HEP Data Analysis}
\label{sec:anomalies:model}

A typical search for new phenomena begins by selecting a target model or class of models, $S$.   Then, simulations of the signal and simulations of the Standard Model (SM) background $B$ are performed.  These simulations are used to inform the construction of a test statistic $\lambda$ that is based on features $x\in\mathbb{R}^N$.  A threshold $c$ is chosen so that $\Pr(\lambda > c|S+B)$ is significantly larger than $\Pr(\lambda > c|B)$.  A combination of simulations and data-driven methods are used to estimate these tail probabilities given the observed data.  A discovery is declared when the data are much more consistent with the $S+B$ hypothesis than the $B$-only hypothesis.  If instead the data are more consistent with the $B$-only hypothesis, then one can set limits on the production cross section of a hypothetical signal.

In this paradigm, models are used in two important ways.  First, models of both $S$ and $B$ are used to choose $\lambda$.  In the absence of nuisance parameters and for a single signal model, the Neyman-Pearson lemma~\cite{neyman1933ix} states that for a fixed probability of rejecting the null hypothesis when it is true (level), the probability for rejecting the null hypothesis when the alternative is true (power) is maximized with the likelihood ratio test statistic.   Typically, $\lambda$ is chosen manually by scanning features that are known to enhance the likelihood ratio.   A growing number of searches use various machine learning methods to strive for optimal performance.   Machine learning methods can also be used to extend beyond a `cut-and-count' approach to achieve an unbinned optimal test statistic~\cite{Nachman:2019dol}.  When a search involves profiling nuisance parameters or composite alternative hypotheses (multiple signal models), there is no uniformly most powerful test statistic.   In these cases, the likelihood ratio is still a reasonable target and there are also a variety of inference-aware methods for achieving optimality~\cite{deCastro:2018mgh,Wunsch:2020iuh}.

Second, models are used to determine the $p$-values $\Pr(\lambda > c|S+B)$ and $\Pr(\lambda > c|B)$.  In some cases, simulations are used to directly estimate these probabilities using Monte Carlo (MC) methods.  In many cases, data are used as part of the density estimation.  The extent to which data can be used depends on assumptions about the background and the signal in the targeted region of phase space.   Even if the $p$-values are ultimately computed entirely from data using parametric or non-parameteric methods, simulations often play a key role in validating the assumptions that justify the data-based procedure or in deriving method non-closure uncertainties. 

\begin{figure}[h!]
\centering
\includegraphics[width=0.95\textwidth]{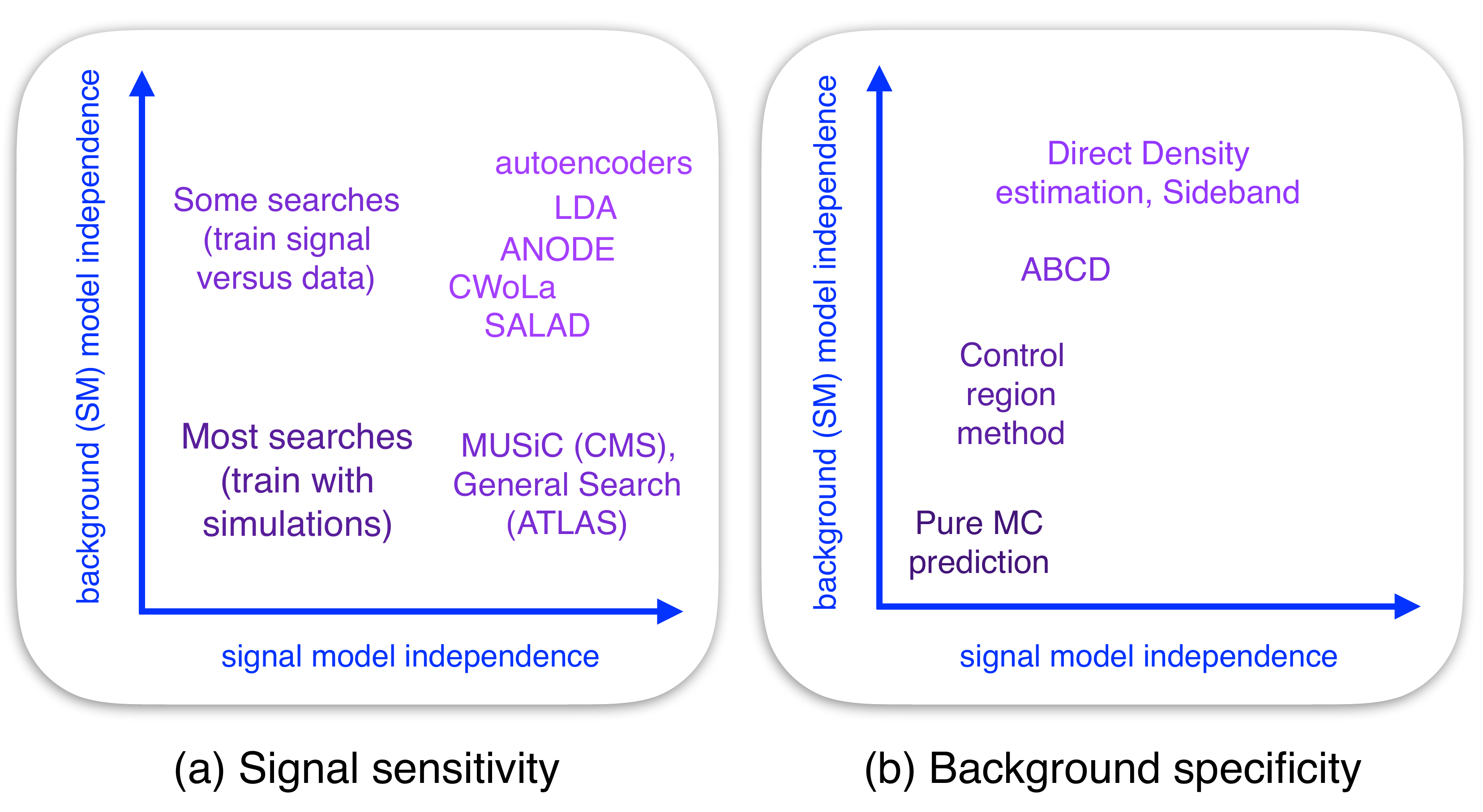}
\caption{A graphic depicting the classification of searches by their dependence on models of the signal and Standard Model (SM) background for both (a) gaining signal sensitivity (choosing the test statistic $\lambda$) and (b) calibrating the background $p$-values.  The Model Unspecific Search for New Physics (MUSiC)~\cite{CMS-PAS-EXO-14-016,CMS-PAS-EXO-10-021,Sirunyan:2020jwk} and General Search~\cite{Aaboud:2018ufy,ATLAS-CONF-2014-006,ATLAS-CONF-2012-107} are results from the CMS and ATLAS Collaborations, respectively.   LDA stands for \textit{Latent Dirichlet Allocation}~\cite{10.5555/944919.944937,Dillon:2019cqt}, ANODE stands for \textit{ANOmaly detection with Density Estimation}~\cite{Nachman:2020lpy}, SALAD stands for \textit{Simulation Assisted Likelihood-free Anomaly Detection}~\cite{Andreassen:2020nkr} and CWoLa stands for \textit{Classification Without Labels}~\cite{Metodiev:2017vrx,Collins:2018epr,Collins:2019jip}.  Direct density estimation is a form of side-banding where the multidimensional feature space density is learned conditional on the resonant feature.  Figure reproduced from Ref.~\cite{Nachman:2020lpy}.
}
\label{fig:schematic}
\end{figure}

One can categorize a search strategy by its $S$- and $B$-dependence for both the signal sensitivity and background specificity, as shown in Fig.~\ref{fig:schematic}.  As described above, most searches can be placed in the lower left part of Fig.~\ref{fig:schematic}(a) (signal sensitivity).   Searches with a clear signal hypothesis that can be accurately simulated, but with a complex or hard-to-simulate background are found in the upper left part of Fig.~\ref{fig:schematic}(a).  The searches depicted in the lower right corner of Fig.~\ref{fig:schematic}(a) will be discussed further in Sec.~\ref{sec:fullysupervised} and the remaining search strategies illustrated in the top right corner of Fig.~\ref{fig:schematic}(a) will be described in Sec.~\ref{sec:unsupervised}, \ref{sec:weaklysupervised}, and~\ref{sec:hybrid}.  The notion of optimality is more complicated for these \textit{anomaly detection} approaches because there is no single signal model hypothesis.  However, one can still consider optimality in the context of a different hypothesis test: data versus background.  Let $p_\text{background}(x|B)$ be the density describing the background and $p_\text{data}(x)$ be the density observed in a signal region.  One can view anomaly detection as positing $p_\text{background}(x|B)$ as the null hypothesis with $p_\text{data}(x)$ as the alternative hypothesis.  By construction, the data are consistent with the alternative hypothesis, but this does not mean that the null hypothesis can be rejected.  As a simple hypothesis test, the Neyman-Pearson lemma guarantees that $p_\text{background} / p_\text{data}$ is the optimal test statistic.   An anomaly detection technique is defined to be \textit{asymptotically optimal} if there is some limit in which it approaches this optimal test statistic.   While many of the techniques in Sec. ~\ref{sec:weaklysupervised} and~\ref{sec:hybrid} are asymptotically optimal, most of the methods presented in Sec.~\ref{sec:unsupervised} are not.

Achieving signal sensitivity is not sufficient to produce a physics result~--~one must also calibrate the background.  Said another way, selecting anomalous events is irrelevant if there is no context to provide information on their strangeness.  Precise background calibration is a key difference in anomaly detection between industry and HEP.  In the former, anomalies are often off-manifold (e.g. a flying elephant) instead of local over-densities (e.g. more elephants than expected at a watering hole) and so it is less important that the background rate be known precisely.   A variety of common methods are highlighted in Fig.~\ref{fig:schematic}(b).  Pure MC estimation is reserved for final states that are relatively simple and precisely known theoretically and experimentally such as pure electroweak processes with only charged leptons~\cite{Aaboud:2017rel}.  Many searches use the \textit{control region method} whereby simulations are calibrated using event selections that are close to the final phase space, but sufficiently far away that the expected signal purity is low.  The requirement on signal purity introduces a dependence on the signal model.  Two `fully data driven' approaches are the ABCD and sideband methods.  In both approaches, the data are used directly to transport predictions from a control region to the signal sensitive region instead of relying on simulation for this extrapolation.  In the ABCD method, two classifiers $f$ and $g$ and two working points $a$ and $b$ are constructed and then four regions called $A,B,C$ and $D$ are defined by $f\lessgtr a$ and $g \lessgtr b$ (for a machine learning version of ABCD, see~\cite{Kasieczka:2020pil,Choi:2020bnf,Aad:2020hzm}).  If $f$ and $g$ are independent, then one can relate the background prediction in the region $f>a$ and $g>b$ to the other three regions.  The ABCD method depends on the background model by verifying independence and on the signal model by verifying low signal contamination in the sidebands.  The sideband method requires knowledge of one feature where the signal should be localized and the background is not localized.  Regions away from the signal are then defined as sidebands and a fit can be used to interpolate the background prediction into the signal region.  Resonant new physics is expected to be localized at the mass of the new particle, so this is a moderately weak assumption.  Relatively simple parametric models are often used for the fit, although machine learning approaches have also been proposed for this purpose~\cite{Frate:2017mai,DiSipio:2019imz,Andreassen:2020nkr,Nachman:2020lpy}.

A core requirement of the sideband method is that the background is not localized where the potential signal should be localized.  While this is generally true inclusively because background processes are often non-resonant, machine learning classifiers can artificially introduce localized features in the background.  This often happens if a classifier is trained to distinguish a resonant signal from the SM background and a feature sensitive to the mass of the new particle is used in the training.  One can simply remove mass-sensitive features from the training, but powerful classifiers can learn the mass indirectly through subtle correlations with other useful features.   A variety of decorrelation techniques exist to solve this problem~\cite{Louppe:2016ylz,Dolen:2016kst,Moult:2017okx,Stevens:2013dya,Shimmin:2017mfk,Bradshaw:2019ipy,ATL-PHYS-PUB-2018-014,DiscoFever,Xia:2018kgd,Englert:2018cfo,Wunsch:2019qbo,Aguilar-Saavedra:2017rzt,Sirunyan:2020lcu,Kitouni:2020xgb,Kasieczka:2020pil}.   In the context of neural networks, one can add terms to the loss function to achieve automatic decorrelation:

\begin{align}
\mathcal{L}(f)=\sum_i \mathcal{L}_\text{classifier}(f(x_i),y_i) + \alpha \, (1-y_i)\,\mathcal{L}_\text{decorrelation}(f(x_i),m_i),
\end{align}

\noindent where $\alpha\geq 0$ is a hyperparameter, $\mathcal{L}_\text{classifier}$ is the usual classifier loss such as binary cross entropy or mean squared error, $f$ is the classifier, $x$ are the features, $y$ are the labels ($y=0$ is background), and $m$ is the feature that needs to be independent from $f$.  In one approach, $\mathcal{L}_\text{decorrelation}$ is itself a neural network~\cite{Shimmin:2017mfk} that is designed to learn $m$ from $f$.  This adversarial method is optimized as a minimax solution whereby the second neural network is as bad as possible when decorrelation is achieved.  Another possibility is for\footnote{Technically, for the distance correlation, this loss is applied at the level of a batch because it requires computing expectation values over pairs of events.} $\mathcal{L}_\text{decorrelation}$ to be a measure of dependence between $f(x)$ and $m$ such as the distance correlation~\cite{DiscoFever}.  The later introduces only one free parameter ($\alpha$) compared with thousands of parameters in adversarial method, but may be less flexible and can require large batches during training.  It is also possible to relax the decorrelation assumption by allowing for a controlled dependence on $m$~\cite{Kitouni:2020xgb}.

In practice, analyses can mix and match strategies from Fig.~\ref{fig:schematic}(a) and Fig.~\ref{fig:schematic}(b).  Some methods are naturally paired, such as those described in Sec.~\ref{sec:weaklysupervised} and~\ref{sec:hybrid}, as they were developed in the context of the sideband method.   Decorrelation techniques that were first introduced for background estimation can also be important for signal sensitivity.  The interplay between independence and signal sensitivity is discussed in Sections~\ref{sec:weaklysupervised} and~\ref{sec:hybrid}.  The remainder of this chapter focuses on machine learning methods for achieving signal sensitivity.

\section{Signal Independent, Background Model Dependent}
\label{sec:fullysupervised}

In some sense, all searches for physics beyond the Standard Model are anomaly detection analyses because anything discovered would be anomalous by definition.   Furthermore, many searches publish `model-independent limits'.  Such limits are signal model-independent only in the cross-section, but are strongly signal model-dependent in the acceptance.    The rest of this chapter will instead focus on searches that do not target a particular signal model, although they often target a class of models such as resonance decays into a particular final state.   Many searches are relatively inclusive and not optimized for a particular signal (e.g. the inclusive dijet search at the LHC~\cite{Aad:2019hjw,Sirunyan:2019vgj}).  While such searches are broadly sensitive, this chapter will not focus on them because they are not actively optimized for sensitivity to new physics aside from general considerations (e.g. avoid large rapidity gaps in the case of dijets).   Signal model-independent searches have a long history in HEP and have been performed by D0~\cite{sleuth,Abbott:2000fb,Abbott:2000gx,Abbott:2001ke}, H1~\cite{Aaron:2008aa,Aktas:2004pz}, ALEPH~\cite{Cranmer:2005zn}, CDF~\cite{Aaltonen:2007dg,Aaltonen:2007ab,Aaltonen:2008vt}, CMS~\cite{CMS-PAS-EXO-14-016,CMS-PAS-EXO-10-021,Sirunyan:2020jwk,Sirunyan:2020jwk}, and ATLAS~\cite{Aaboud:2018ufy,ATLAS-CONF-2014-006,ATLAS-CONF-2012-107}.   The general strategy in these analyses is to directly compare data with simulation in a large number of exclusive final states (bins).   

Recent proposals extend this methodology to be unbinned by using nearest neighbors~\cite{DeSimone:2018efk}, cluster similarity~\cite{1809.02977} and neural networks~\cite{DAgnolo:2018cun,DAgnolo:2019vbw}.  Another innovation in Ref.~\cite{DAgnolo:2018cun} is the introduction of a new loss function that leads the neural network to learn the log likelihood ratio directly:

\begin{align}
\mathcal{L}(f)=\sum_i (1-y_i)(e^{f(x_i)}-1)-y_i\,f(x_i).
\end{align}

\noindent In the asymptotic limit (sufficient training data and network/training flexibility), one could then interpret the neural network output as log likelihood ratio which has analytic formulae for computing $p$-values~\cite{wilks1938,Cowan:2010js,10.2307/1990256}.  Interpreting the output directly as a test statistic / likelihood ratio estimator has also been used for a variety of other studies related to simulation-based inference and domain adaptation~\cite{Andreassen:2019nnm,Stoye:2018ovl,Hollingsworth:2020kjg,Brehmer:2018kdj,Brehmer:2018eca,Brehmer:2019xox,Brehmer:2018hga,Cranmer:2015bka,Badiali:2020wal,Andreassen:2020nkr,Andreassen:2019cjw,2010.03569}.

The approaches mentioned above are nearly signal model independent.  The only signal model dependence is in the selection of the targeted phase space and analysis features.  This is particularly beneficial for non-resonant new physics\footnote{This may not apply to cases of strong signal-background interference.}, where there are fewer methods available.  However, there is a strong dependence on the background model.  Any significant sources of mis-modeling in the background will be flagged as anomalies.  Therefore, potential signals need to be larger than these mis-modelings and the false positive rate must be calibrated to account for the expected deviations between data and simulation.

\section{Supervised Approaches}
\label{sec:supervised}

The remainder of this chapter focuses on searches that do not use differences between background simulations and data to directly achieve signal sensitivity.   Many of the following methods still heavily rely on background simulations for the most natural background calibration method, but that is not discussed further here.  In general, techniques can be classified based on their level of \textit{supervision}.  Fully supervised approaches use data with labels of signal and background for the $y_i$ in their loss functions.  Unsupervised approaches have no per-instance labels and must resort to other methods of identifying structure in data.  A variety of methods called \textit{weakly supervised} are based on incomplete label information and will be introduced in Sec.~\ref{sec:weaklysupervised}.

One strategy for supervised anomaly detection is to train with a signal simulation that includes a variety of individual signal processes.   This idea was explored in one of the first modern machine learning based anomaly detection procedures called \textit{anti-QCD tagging}~\cite{Aguilar-Saavedra:2017rzt} (see also~\cite{Aguilar-Saavedra:2020uhm}).  Instead of using a discrete set of signal models~\cite{Khosa:2020qrz}, a signal dataset was generated using flat matrix elements.  This means that there is no special energy scale in the problem and there is broad coverage for all of the kinematic configurations that can occur with a certain number of final state objects.  In this way, no particular signal masses were preferred, but the classifier was able to learn generic features of a large class of signal models such as the number of prongs within a jet.  A generic feature of anomaly detection that was well-characterized in Ref.~\cite{Aguilar-Saavedra:2017rzt} is that these approaches are less sensitive than dedicated searches for targeted signal models.  However, anomaly detection methods can be more sensitive than dedicated searches for non-targeted signal models.

\section{Unsupervised Approaches}
\label{sec:unsupervised}

\begin{figure}[h!]
\centering
\includegraphics[width=0.95\textwidth]{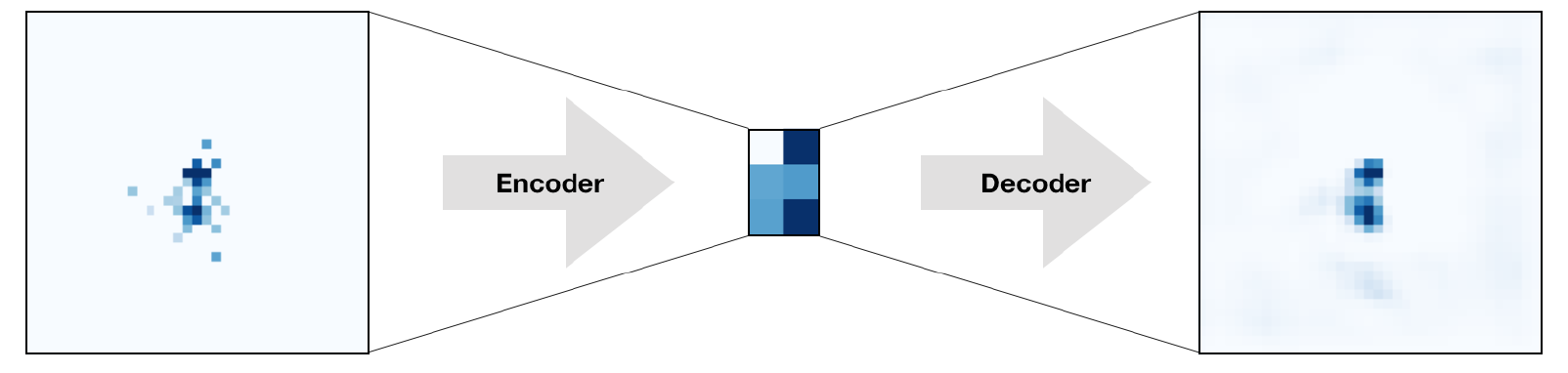}
\includegraphics[width=0.95\textwidth]{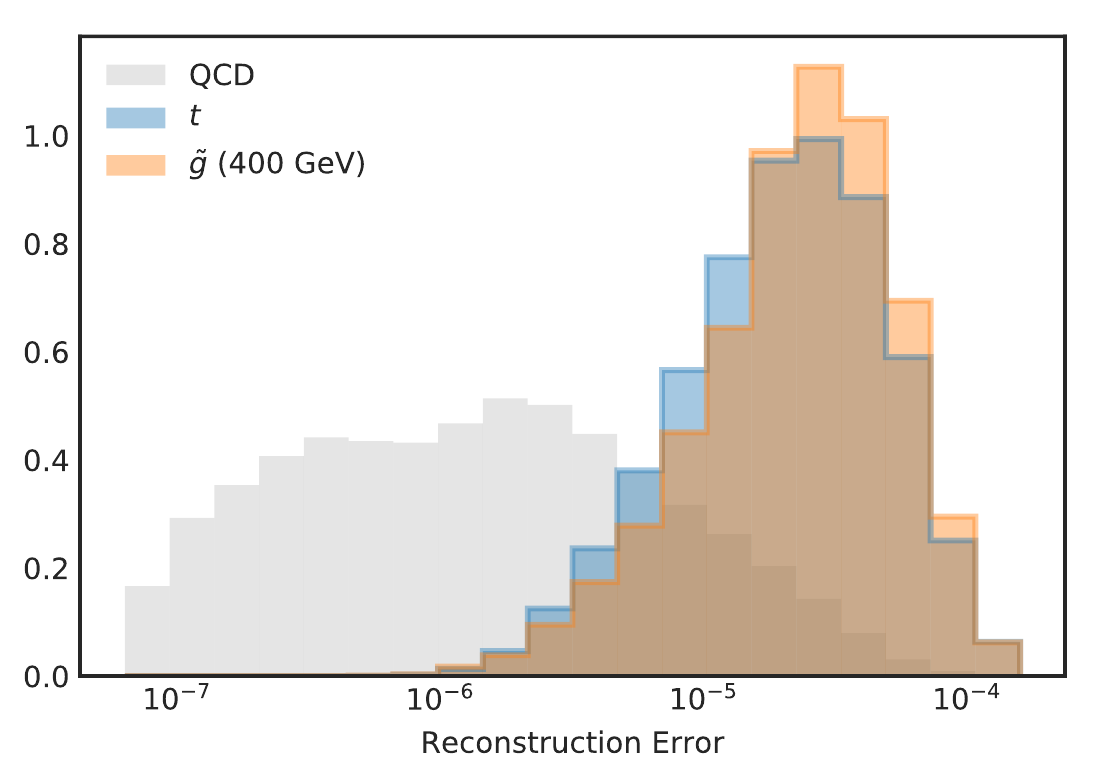}
\caption{An illustration of a vanilla autoencoder-based anomaly detection strategy.  The top diagram illustrates how the image of a jet is compressed and then restored with two neural networks.  The bottom plot is the reconstruction error $|f(g(x))-x|^2$, which is higher for processes that were not used or at least minimally present in the training of the autoencoder.  Figure reproduced from Ref.~\cite{Farina:2018fyg}.
}
\label{fig:AE}
\end{figure}

Unsupervised methods do not use per-instance labels.   The strategy of these approaches is to identify events that are unlike typical background events.  One way to do this would be to learn the density $p_\text{background}(x)$ and then consider events with $p_\text{background}(x)\ll 1$.   Learning high-dimensional densities is difficult and so this has not yet been studied\footnote{Density estimation has been studied in the context of likelihood ratio estimation~\cite{Nachman:2020lpy}.}.  Instead, the approaches that have been studied so far use compression methods (see also \textit{representation learning}~\cite{10.1109/TPAMI.2013.50}).  Two networks $f$ and $g$ encode and decode data, respectively.   These networks are trained to be near inverses of each other: $f(g(x))\approx x$.   The target of $f$ is regularized so that the compression is lossy.   The idea is that events with $f(g(x))\approx x$ should be common and thus ignored while events with $f(g(x))$ far from $x$ are anomaly candidates.  This strategy has been studied in the context of autoencoders (AE) / variational autoencoders (VAE~\cite{kingma2014autoencoding,Kingma2019})~\cite{Farina:2018fyg,Heimel:2018mkt,Roy:2019jae,Blance:2019ibf,Hajer:2018kqm,Cerri:2018anq,Cheng:2020dal} and generative adversarial networks (GANs~\cite{Goodfellow:2014:GAN:2969033.2969125})~\cite{Knapp:2020dde}.  Figure~\ref{fig:AE} illustrates the method for vanilla autoencoders.  The lossy compression is achieved by limiting the expressivity of $f$ and the dimensionality of $f(x)$.  Both $f$ and $g$ are trained at the same time when minimizing the loss $|f(g(x))-x|^2$.  A classifier is created using the loss (`reconstruction error').  Ideally, typical events should have a lower reconstruction loss than events that were not used or at least minimally present in the training of the autoencoder.  This is explicitly demonstrated in the bottom plot of Figure~\ref{fig:AE}, where the more typical generic QCD jet processes have a much lower reconstruction loss than top quark pair production or gluino production.  Other strategies for regulating the latent space are possible, such as requiring it to be of a particular form.  Variational autoencoders (VAEs) use this strategy, where the latent space is typically required to be a multivariate Gaussian distribution.  The anomaly detection based on a GAN presented in Ref.~\cite{Knapp:2020dde} is conceptually similar to the way AEs are used for anomaly detection, but the model is trained using a bidrectional GAN~\cite{donahue2016adversarial} instead of a variational autoencoder setup.  Another unsupervised approach using a clustering method was proposed in Ref.~\cite{Mikuni:2020qds}.

A challenge with unsupervised approaches is that the signal may not occupy regions of low $p_\text{background}$.  If instead, the signal is an overdensity in a region of relatively high background probability density, then it may be well-reconstructed by compression methods.  Additionally, a feature of some compression methods is that high reconstruction loss may not necessarily correspond to low $p_\text{background}$.  For example, consider a vanilla autoencoder trained to compress two dimensions into one dimension.  Further suppose that the background is a bivariate Gaussian.   A limited capacity network may simply learn $f(x_0,x_1)=\rho\frac{\sigma_1}{\sigma_0}(x_0-\mu_0)+\mu_1$, where $\mu_i,\sigma_i$ are the mean and standard deviation of the $i^\text{th}$ direction and $\rho$ is the correlation coefficient.  This function is the minimum variance unbiased estimator of $x_1$ given $x_0$.   If the signal is also a Gaussian with mean far from the line defined by $f(x)$, then it will have a poor reconstruction error.  However, the signal could also be on the line, but from the $(\mu_0,\mu_1)$.  This would correspond to a very low $p_\text{background}$, but also a low reconstruction error.  Despite these challenges, AEs are popular in HEP and beyond (see Ref.~\cite{PIMENTEL2014215} and papers that cite it) and are sufficiently generic that they may be complementary to the approaches described in the next sections.

\section{Weak Supervision and Topic Modeling}
\label{sec:weaklysupervised}

\begin{figure}[h!]
\centering
\includegraphics[width=0.45\textwidth]{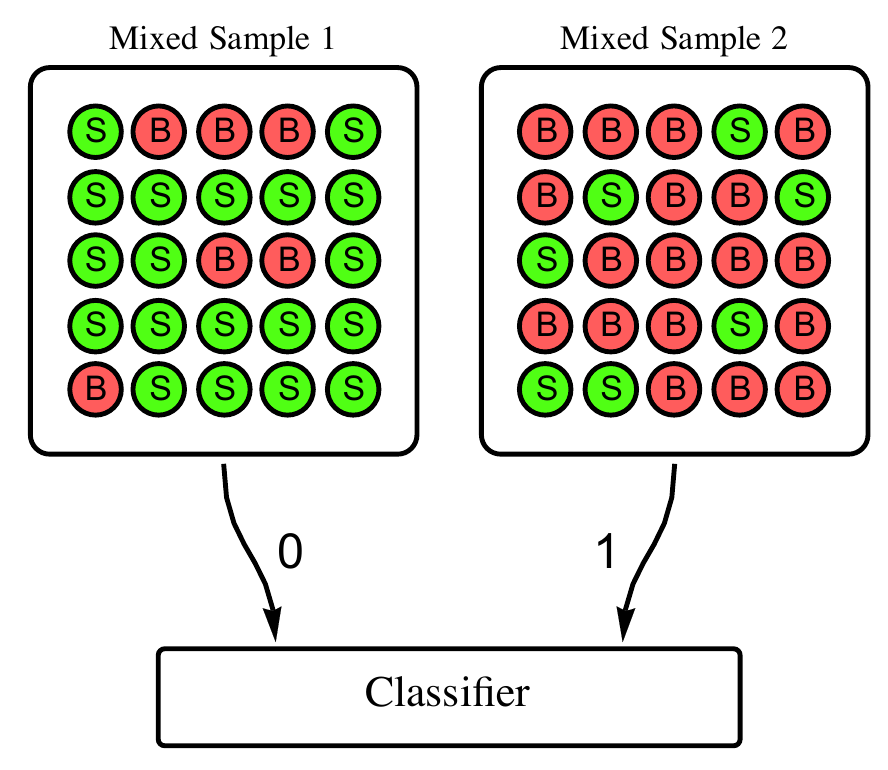}
\includegraphics[width=0.45\textwidth]{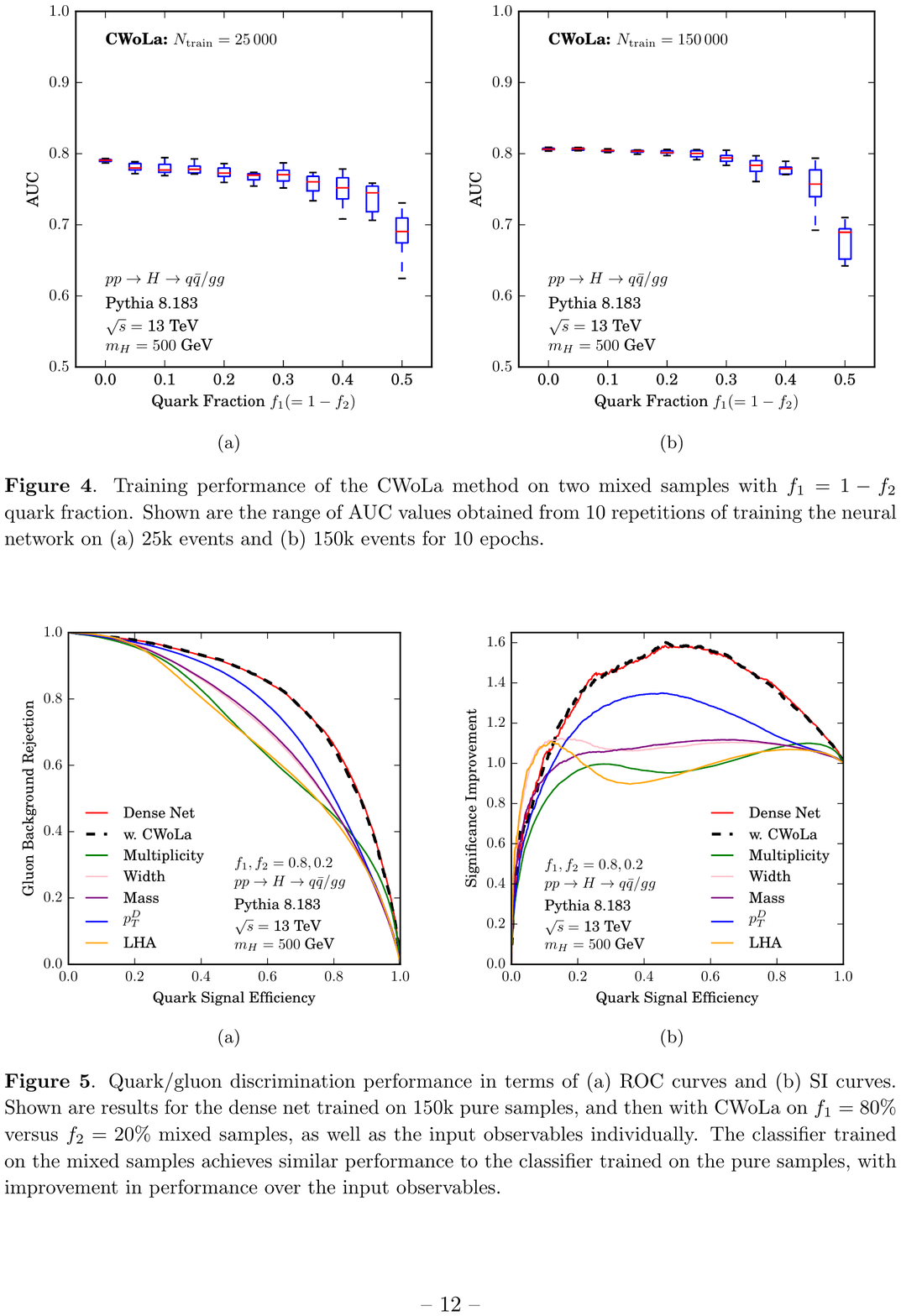}
\caption{The left diagram illustrates the setup of the CWoLa method and the right plot demonstrates that the weakly supervised CWoLa can achieve the same performance as a fully supervised classifier trained with the same features.  In the right plot, a better classifier would be up and to the right (gluon jet rejection is one minus the gluon jet efficiency).  Figure reproduced from Ref.~\cite{Metodiev:2017vrx}.
}
\label{fig:cwola}
\end{figure}

One challenge with the unsupervised methods in the previous section is that they do not explicitly use the (potential) presence of signal in the data.  This section will introduce methods that make use of the presence of the signal by using supervised learning without explicit per instance labels.   One setting where this is possible is the case of mixed samples.  Suppose that there are two sets of data $\mathcal{M}_0$ and $\mathcal{M}_1$, each of which is a mixture of $S$ and $B$.  If the per-instance labels are known, then one can train a fully supervised classifier.  However, if these sets are from real events, then per-instance labels are unknown.  A variety of weakly supervised methods have been developed for this setting~\cite{Metodiev:2017vrx,Dery:2017fap,Komiske:2018oaa,Cohen:2017exh}.  The first of these posited that the fractions $t$ of signal in each dataset are known.  Then, one can train a weakly supervised classifier:

\begin{align}
f_\text{weak}=\text{argmin}_{f':\mathbb{R}^n\rightarrow [0,1]}\mathcal{L}\left(\sum_{i=1}^N\frac{f'(x_i)}{N}-t\right),
\end{align}
which should be contrasted with the fully supervised case:
\begin{align}
f_\text{supervised}=\text{argmin}_{f':\mathbb{R}^n\rightarrow [0,1]}\sum_{i=1}^N\mathcal{L}\left(f'(x_i)-y_i\right),
\end{align}
where in both cases, $\mathcal{L}$ is a loss function such as mean squared error.  This modification to the loss is effective, but (a) requires significant modification to the learning setup (learn on average instead of per-instance) and (b) requires the fractions to be known\footnote{The label fractions need not be known exactly for optimal performance~\cite{Cohen:2017exh}.}.  An alternative approach is the Classification without labels (CWoLa)~\cite{Metodiev:2017vrx} framework.  In this setup, one \textit{assigns} labels to each event and then performs supervised learning with the assigned labels.  In particular, all events from $\mathcal{M}_0$ are labeled $0$ and all the events in $\mathcal{M}_1$ are labeled 1.  This is illustrated in the left part of Fig.~\ref{fig:cwola}.  One can show that if the CWoLa classifier is optimal for distinguishing $\mathcal{M}_0$ from $\mathcal{M}_1$, then it will be also optimal at distinguishing\footnote{If $0\leftrightarrow 1$, then one may learn to anti-tag the signal.  This simply then requires that one know which of $\mathcal{M}_0$ or $\mathcal{M}_1$ is expected to have a higher signal proportion.} $S$ from $B$.  The right plot of Fig.~\ref{fig:cwola} shows that the CWoLa classifier achieves the same performance as a fully supervised classifier on a particular quark-jet versus gluon-jet classification task.  This is particularly powerful because the fully supervised classifier has per instance labels $y_i$ while the CWoLa classifier has far less information.   Before proceeding, it is important to consider the assumptions of the CWoLa method.  Most importantly, this approach only works if the differences between $p(S|\mathcal{M}_0)$ and $p(S|\mathcal{M}_1)$ (similarly for the background) are small compared to the difference between $p(S|\mathcal{M}_0)$ and $p(B|\mathcal{M}_0)$.  In other words, the signal events from $\mathcal{M}_0$ and $\mathcal{M}_1$ should be statistically identical (and the same for the background).  The only other requirement is that the signal fractions should not be the same.  The effective statistics available to learn scale like $n_\text{signal}(y_0-y_1)$ so the closer the fractions $y_0$ and $y_1$ are to each other, the worse the classification performance will be.  In the limit $y_0\rightarrow 0$ and $y_1\rightarrow 1$, CWoLa simply approaches fully supervised classification.  Note that in order to make the performance curve in Fig.~\ref{fig:cwola}, one needs at least a small set of labeled examples or the class fractions.  If one does not need to know the actual performance, this is not necessary.

A variety of related concepts have been introduced for similar purposes.  The sPlot~\cite{Pivk:2004ty,Borisyak:2019vbz} method learns to decompose a dataset into its constituent processes without per-instance labels, but it requires knowing $p(x|S)$ and $p(x|B)$ for at least a subset of the features.  The topic modeling introduced in Ref.~\cite{Metodiev:2018ftz} and further studied in Ref.~\cite{Komiske:2018vkc,Metodiev:2018ftz,Aad:2019onw,Alvarez:2019knh} relaxes this requirement by extracting information directly from data using extreme regions of phase space that are nearly pure $S$ or $B$.   Another variation that seeks to solve the challenge of different classes sharing common features is the Latent Dirichlet Allocation (LDA)~\cite{10.5555/944919.944937} approach to mixed-membership models in Ref.~\cite{Dillon:2020quc,Dillon:2019cqt}.

\begin{figure}[h!]
\centering
\includegraphics[width=0.6\textwidth]{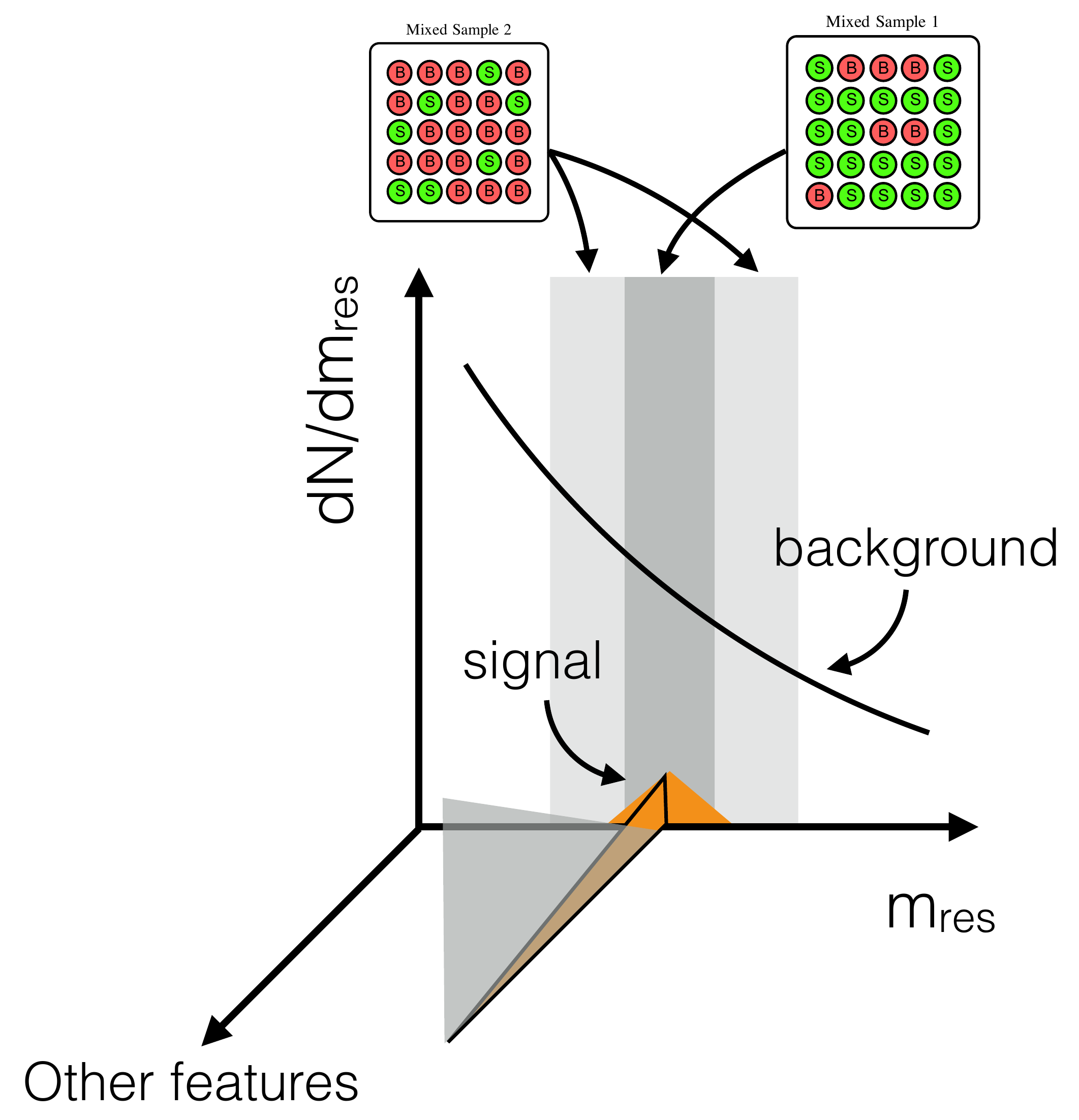}
\includegraphics[width=0.95\textwidth]{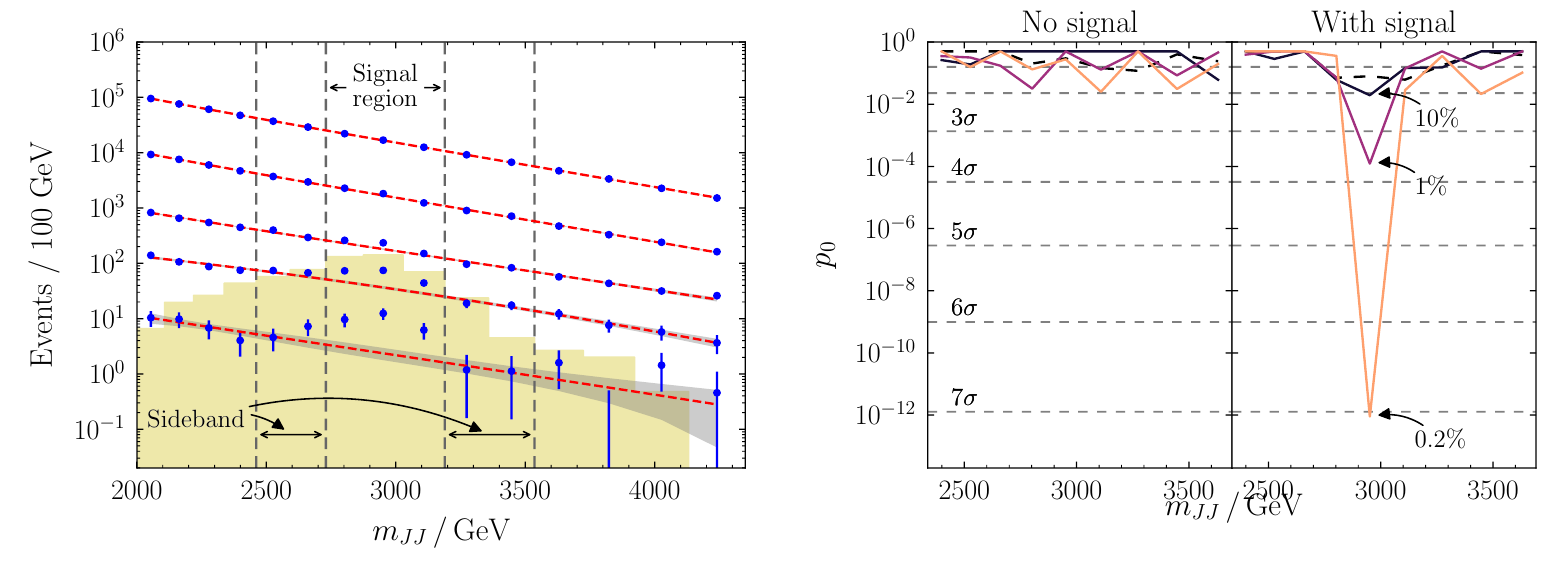}
\caption{Top: a schematic diagram illustrating the CWoLa setup for the anomaly detection task.  Bottom: figures reproduced from Ref.~\cite{Collins:2018epr,Collins:2019jip} that show the application of the CWoLa hunting method to the dijet search at the LHC.  The bottom left plot is the dijet invariant mass with the different lines corresponding to increasingly tighter thresholds on the CWoLa classifier.  The injected signal is many order of magnitude below the background.  The bottom right plot shows the extracted $p$-value for a scan in the signal region location.  Without any injected signal, the $p$-values are consistent with uniform while when there is signal injected, the initially 1.5$\sigma$ excess is automatically enhanced to a 7$\sigma$ potential discovery.
}
\label{fig:cwolahunting}
\end{figure}

There are many ways to combine the weakly supervised methods described above with anomaly detection.  One approach combines CWoLa with a bump hunt in a feature $m_\text{res}$ as illustrated in Fig.~\ref{fig:cwolahunting}.  In particular, two mixed samples are constructed by using a signal region around a hypothetical resonance and a sideband region.  The sideband region should be as close as possible to the signal region in order to ensure that the background is nearly the same in the two mixed samples.  The other features used for training the CWoLa classifier need to be nearly independent of $m_\text{res}$.  The bottom plots in Fig.~\ref{fig:cwolahunting} illustrates the performance of this `CWoLa hunting' method to a dijet search at the LHC using the dijet mass as $m_\text{res}$ and other jet substructure features to train the CWoLa classifier.   In the absence of an injected signal, the $p$-values are consistent with uniform while when there is signal injected, the initially 1.5$\sigma$ excess is automatically enhanced to a 7$\sigma$ potential discovery.  In order to make the most use of the data and to avoid a large trials factor, this search involves a complex $k$-fold cross-training procedure.   The look elsewhere effect would be significant if the same data were used for training and applying the CWoLa classifier as local fluctuations would be amplified.  One way around this is to divide the data in half and train one one part and test on the other.  The local fluctuations in both halves are uncorrelated and thus there is no additional trails factor aside from the one in the scan of $m_\text{res}$.  This procedure is extended to $k$-fold in Ref.~\cite{Collins:2018epr,Collins:2019jip} to avoid using only half of the data.

\section{Hybrid Approaches}
\label{sec:hybrid}

\begin{figure}[h!]
\centering
\includegraphics[width=0.9\textwidth]{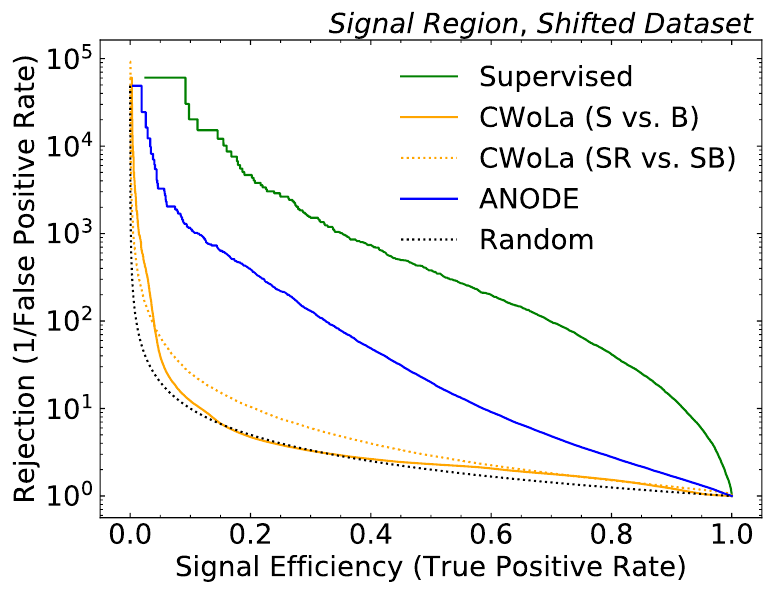}
\caption{Top: the performance of CWoLa and ANODE classifiers on a dataset with artificial correlations.  The CWoLa classifier learns a non-trivial function for signal region versus sideband region and as a result has only random performance on the signal.  In contrast, by learning the numerator and denominator of the likelihood ratio separately, ANODE is more robust to such correlations (but still worse than the dedicated fully supervised tagger).  Figure reproduced from Ref.~\cite{Nachman:2020lpy}.  
}
\label{fig:hybrid}
\end{figure}

One of the main limitations of the CWoLa hunting method is that the features used for training the CWoLa classifier must be nearly independent of $m_\text{res}$.  One hybrid solution proposed to mitigate this problem is ANODE~\cite{Nachman:2020lpy}.   In this method, one first learns $p_\text{sideband}(x|m_\text{res})$ in the sideband regions and then $p_\text{signal region}(x|m)$ from the signal region.   The ratio of the interpolated $p_\text{sideband}(x|m_\text{res})$ to $p_\text{signal region}(x|m)$ serves as an asymptotically optimal likelihood ratio estimator.  There have been significant advances in direct density estimation and the demonstration of ANODE in Ref.~\cite{Nachman:2020lpy} makes use of masked autoregressive flows~\cite{NIPS2017_6828}, a type of normalizing flow~\cite{pmlr-v37-rezende15}.  Density learning is unsupervised because there are no labels and one typically learns $p$ via a maximum likelihood loss function.  The main challenge in direct density estimation is that one needs a neural network that integrates to unity.  In the normalizing flows setting, this is accomplished by starting with a known density (often a Gaussian) and then applying a series of variable changes with tractable Jacobians.   In ANODE, one never compares signal region and sideband region directly and so it is more robust to correlations between $x$ and $m_\text{res}$; in fact, $m_\text{res}$ can be one of the dimensions of $x$.  This is illustrated in Fig.~\ref{fig:hybrid}, which shows that when correlations are artificially introduced to spoil the CWoLa hunting approach, ANODE is still able to provide signal sensitivity.   With advances in neural density estimation (see e.g. Neural Autoregressive Flows \cite{DBLP:journals/corr/abs-1804-00779} and Neural Spline Flows \cite{durkan2019neural}), it is likely that methods like ANODE will continue to improve.  A key difference for HEP compared to industrial applications is that one needs quantitatively precise density estimation - it is not good enough to produce examples that qualitatively look realistic (as in the case of cat and celebrity pictures).

Another hybrid solution is Simulation Assisted Likelihood-free Anomaly Detection (SALAD)~\cite{Andreassen:2020nkr} (see also SA-CWoLa~\cite{2009.02205}).  The motivation of this method is that while simulations are only an approximation to nature, they do encode significant physics information that it would be useful to incorporate into a classifier.   To this end, a parameterized reweighting function $w(x|m)\approx p_\text{data}(x|m)/p_\text{simulation}(x|m)$ is learned in sidebands and then interpolated to the signal region~\cite{Andreassen:2019nnm}.  Any classifier can then be used to distinguish the reweighted simulation in the signal region from the data in that region.   If the reweighting is effective, the result should not depend on the simulation.  At the same time, the closer the initial simulation is to the data, the less reliant the method is on an effective learning and interpolation for $w(x|m)$.

A third hybrid method is Tag N' Train (TNT)~\cite{Amram:2020ykb}, which combined autoencoders with CWoLa.  The motivation is that in the original CWoLa hunting method, the two mixed samples are nearly 100\% of a single class (background) so as a first step, a weak classifier (using an autoencoder - see Sec.~\ref{sec:unsupervised}) slightly purifies the samples before using CWoLa to train a more powerful classifier.   As discussed in Sec.~\ref{sec:weaklysupervised}, the purer the samples, the larger the effective statistics for the CWoLa training and thus the more powerful it will be as a classifier.

It is likely that no one method will be able to cover every scenario and so a diversity of approaches will be needed to ensure broad coverage.  Mixing different approaches may provide further advantages over single algorithms using either entirely supervised or entirely un/weakly supervised methods.

\section{Results with Collider Data}
\label{sec:results}

\begin{figure}[h!]
\centering
\includegraphics[width=0.7\textwidth]{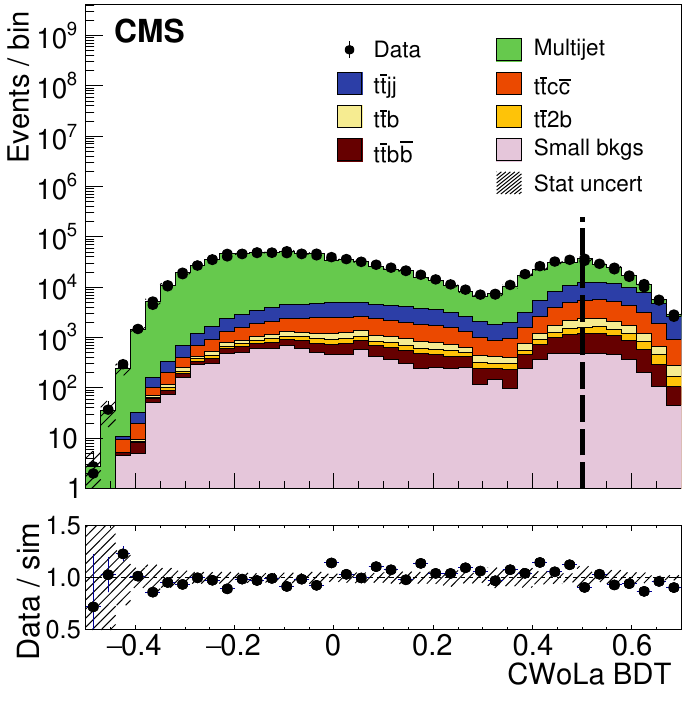}
\caption{A CWoLa classifier (Boosted Decision Tree) used to purify the $t\bar{t}+b\bar{b}$ final state by the CMS Collaboration.  Figures reproduced from Ref.~\cite{Sirunyan:2019jud}.
}
\label{fig:cwolaCMS}
\end{figure}

This chapter will close with a highlight of two recent results from the CMS and ATLAS Collaborations.  Figure~\ref{fig:cwolaCMS} presents the first application of CWoLa\footnote{See Ref.~\cite{Aad:2019onw} for the first measurement to study the related idea of topic modeling.} in a physics analysis.  The CWoLa classifier is used to distinguish generic multijet events from $t\bar{t}$ production and uses two regions with $\mathcal{O}(10\%)$ signal purity.  The ultimate goal of this analysis is to study the $t\bar{t}+b\bar{b}$ process that is an inherently interesting probe of the strong force and is also an important process for Higgs physics in the $t\bar{t}+h$ final state.  Multijet backgrounds are exceptionally difficult to simulate accurately and the CWoLa classifier provides a solution to train directly from data.  While not an anomaly detection search, this analysis demonstrates the feasibility of learning directly from unlabeled data.  

\begin{figure}[h!]
\centering
\includegraphics[width=0.45\textwidth]{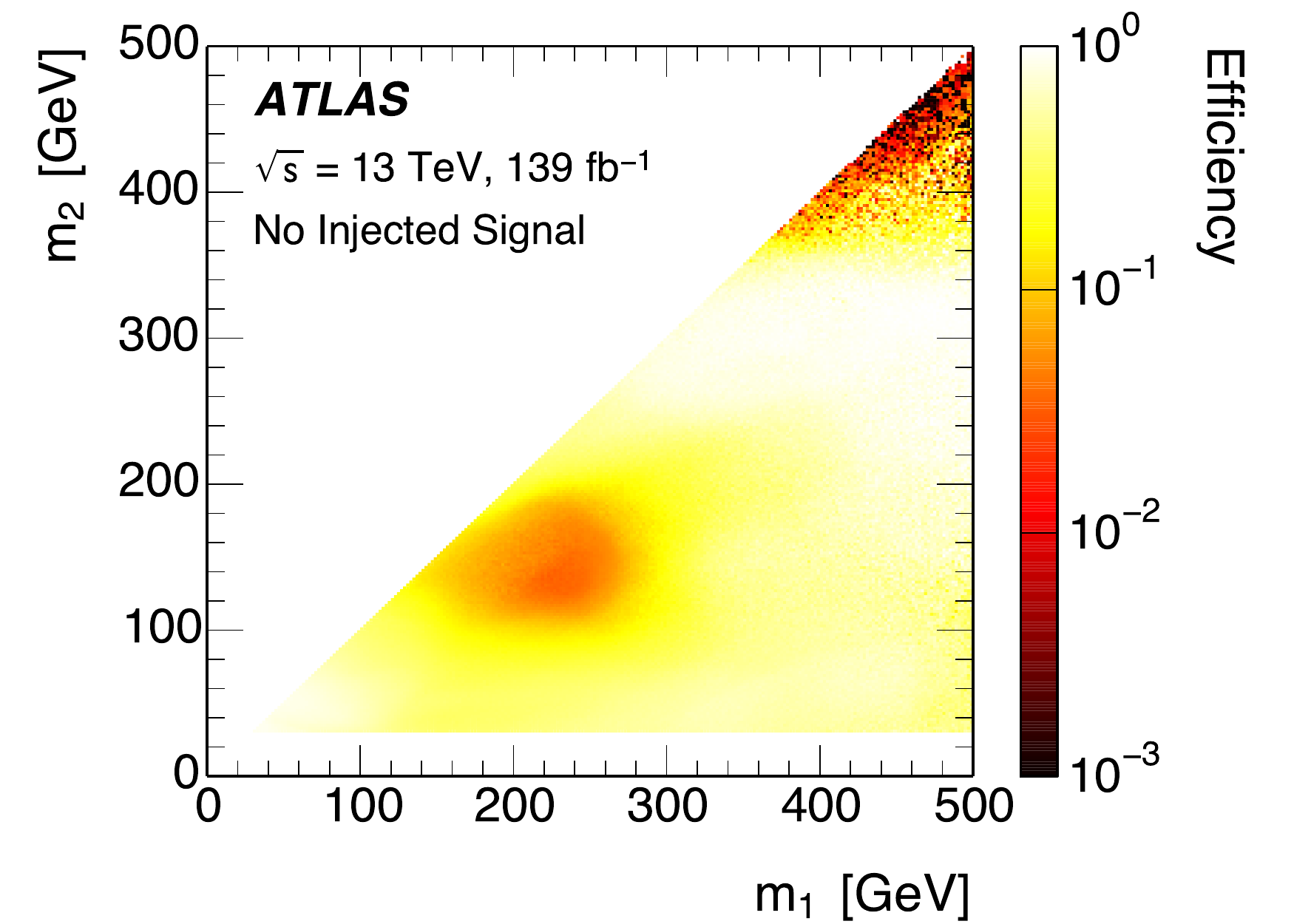}\includegraphics[width=0.45\textwidth]{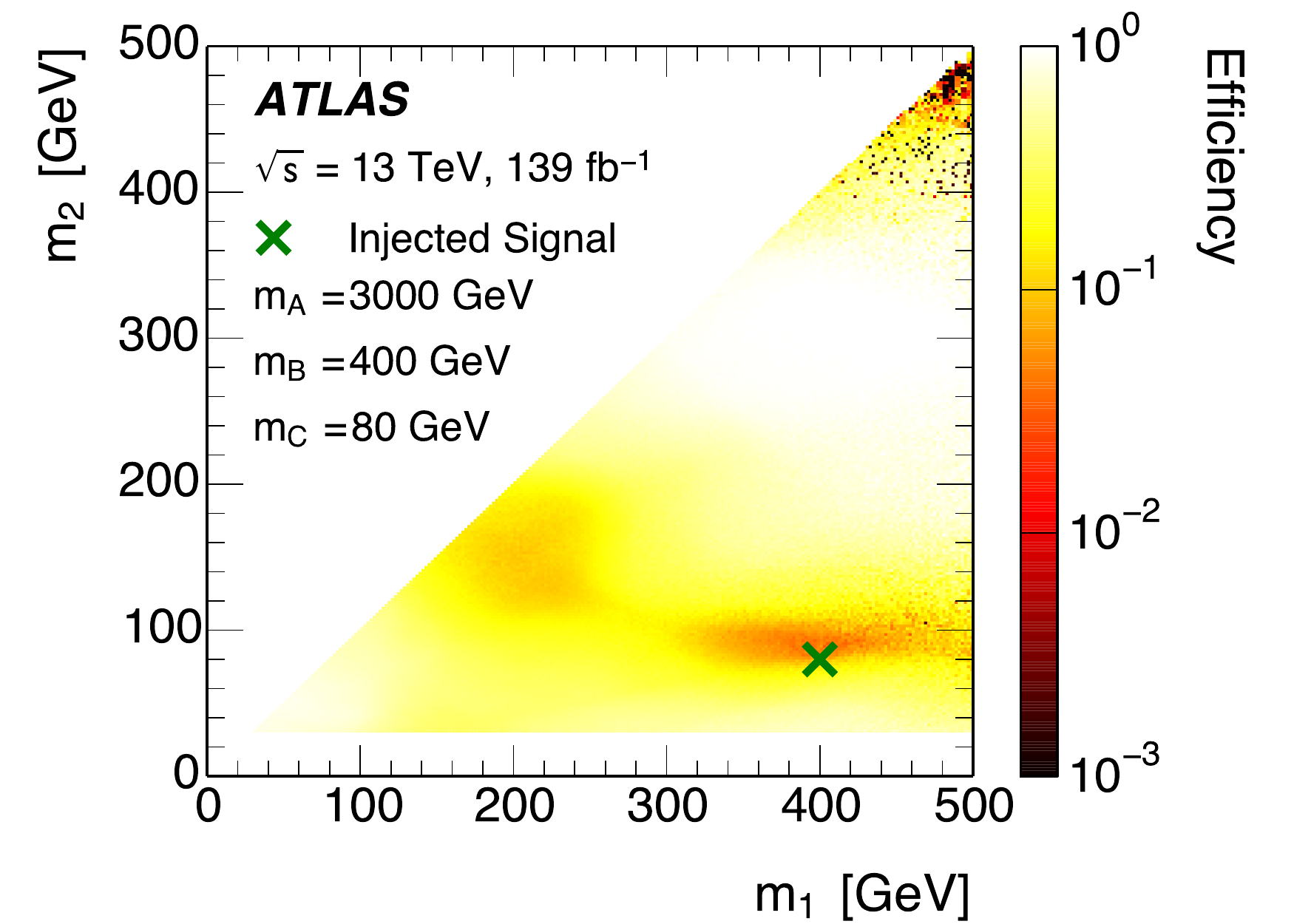}
\includegraphics[width=0.95\textwidth]{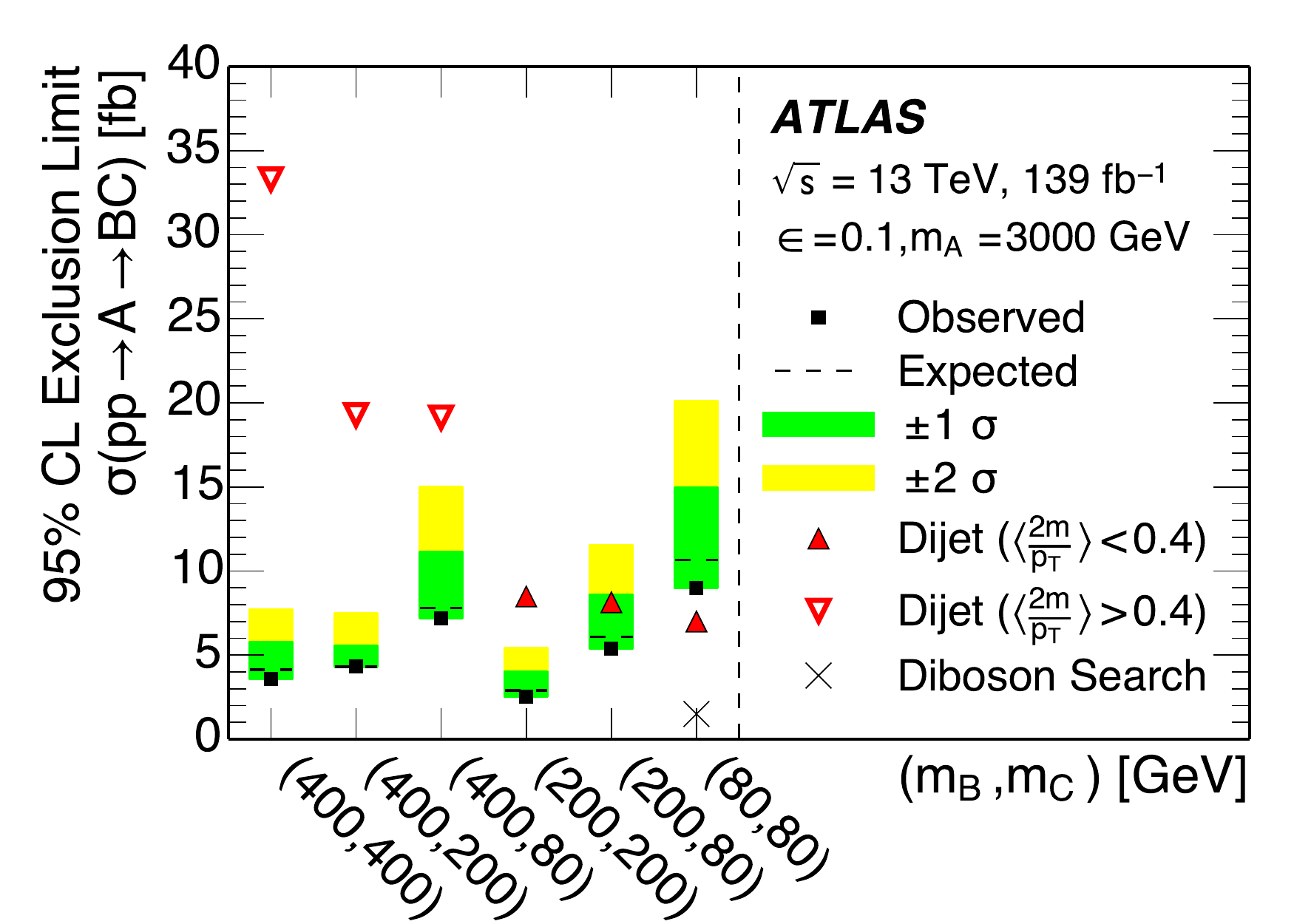}
\caption{Plots from the recent CWoLa hunting search performed by the ATLAS Collaboration~\cite{Aad:2020cws}.  The top two plots show the CWoLa classifier output in one particular signal region on the left with no injected signal and with an injected signal with masses indicated by the cross in the right plot.  The CWoLa classifier is able to automatically identify the signal-like region.  While there was no significant evidence for new particles in this first search, the ATLAS Collaboration has set limits on the production cross section for particular models.
}
\label{fig:cwolahuntingatlas}
\end{figure}

The first machine learning anomaly detection result with data is presented in Fig.~\ref{fig:cwolahuntingatlas} from the ATLAS Collaboration.  This search made use of the CWoLa hunting approach described in Sec.~\ref{sec:weaklysupervised} and targets a dijet final state.  This final state is complex and challenging to model, so data-driven methods are critical.  When the jets are due to hadronically decaying particles that are much lighter than the target resonance mass, their decay products will be collimated and fully contained inside single jets.  The substructure of these jets can be exploited to enhance the signal sensitivity. 

As the first search of its kind, a limited feature space (two-dimensional) was used to help with the technical and sociological integration of this new methodology into the experimental program.  With a two-dimensional feature space, the classifier output can be directly visualized as an image.  Examples of these images are shown in the top plots of Fig.~\ref{fig:cwolahuntingatlas}, where the automatic identification of an injected signal is demonstrated.   The bottom plot of Fig.~\ref{fig:cwolahuntingatlas} shows that for particular models, the CWoLa hunting analysis is able to set the strongest limits for particular models.  One of the most interesting challenges that is like no other analysis is that the event selection depends on the data.  This means that for every injected signal model, for every signal strength, the entire CWoLa hunting procedure has to be rerun.  Aside from making it challenging to recast this analysis, a large number of neural networks ($\mathcal{O}(10^4)$) must be trained to produce Fig.~\ref{fig:cwolahuntingatlas}.   

While there is no evidence for new particles using machine learning methods, this is only the beginning of an expanding program that will produce exciting physics results for many years to come.

\section{Conclusions and Outlook}
\label{sec:outlook}

Given the current lack of convincing evidence for new fundamental structure from HEP experiments, it is critical that the program of dedicated searches be complemented with more model agnostic methods.   The methods presented in this chapter represent a snapshot\footnote{See Ref.~\cite{livingreview} for a more updated list of papers in this area.} of the rapidly developing area of machine learning for anomaly detection in HEP.   

To help catalyze new ideas in anomaly detection for HEP\footnote{A parallel effort under development is described in Ref.~\cite{Brooijmans:2020yij}.  This dataset is planned to be a concoction of SM processes and uses high-level objects instead of low-level hadrons as in the LHC Olympics.  While the current plan does not include multiple generators to emulate `data' and `simulation', the large number of events and heterogeneous composition of physics processes offers an interesting complement to the LHC Olympics.}, the LHC Olympics 2020 was developed~\cite{lhco}.  This data challenge sets up a realistic environment where there is `simulation' and `data', where both are generated on a computer, but possibly from different physics programs.  Furthermore, the `data' comes without labels and may contain some amount of signal.  A variety of methods have been deployed to these datasets, exposing interesting features of the various procedures and helping to prepare for analysis with real data.  With the first results from collider data presented in Sec.~\ref{sec:results}, the field enters a new era where methods must be adapted and modified to meet the needs of real data and new methods must be developed to ensure broad coverage.

Addressing the conceptual, computational, and other challenges associated with the growing field of machine learning-assisted anomaly detection in HEP will make for an exciting research programs in the years ahead.

\section*{\label{sec::acknowledgments}Acknowledgments}

This work was supported by the U.S.~Department of Energy, Office of Science under contract DE-AC02-05CH11231.  I thank Kees Benkendorfer, Jack Collins, Aviv Cukierman, Gregor Kasieczka, Luc Le Pottier, Pablo Mart\'{i}n, and David Shih for countless discussions about anomaly detection that have contributed to the framing presented in this chapter.  Furthermore, I thank Gregor Kasieczka for detailed comments on the manuscript.  I also thank Anders Andreassen, Patrick Komiske, Eric Metodiev, Matt Schwartz, and Jesse Thaler for many helpful discussions about learning from mixed samples.


\bibliographystyle{jhep}
\bibliography{myrefs}

\end{document}